\documentclass[12pt]{article}
\begin{document}
\title{Comments on the Mass of the Photon}
\author{B.G. Sidharth\\
International Institute for Applicable Mathematics \& Information Sciences\\
Hyderabad (India) \& Udine (Italy)\\
B.M. Birla Science Centre, Adarsh Nagar, Hyderabad - 500 063 (India)}
\date{}
\maketitle
\begin{abstract}
De Broglie believed that the photon has a mass, a view shared by a few others. Quite recently, the author has argued that the photon has a mass which is consistent with the latest experimental limits. In the present paper we point out that there is experimental evidence for this mass and also give a theoretical demonstration of the photon mass.
\end{abstract}
\section{Introduction}
As is well known the concept of the photon grew out of the work of Planck and Einstein, though its earliest origin was in Newton's Corpuscular Theory. Thereafter the photon got integrated into twentieth century physics, be it Classical or Quantum. Though it is considered to be a massless particle of spin 1 and 2 helicity states (as proved later for any massless particle with spin by Wigner), it is interesting to note that there had been different dissenting views.\\
De Broglie himself \cite{debroglie} believed that the photon has a mass, a view shared by a few others as well. An apparent objection to this view has been that a photon mass would be incompatible with Special Relativity. However it is interesting to note that nowhere in twentieth century physics has it been proved that the photon has no mass \cite{deser}. It would be correct to say that there are a number of experimental limits to the mass of the photon. These limits have become more and more precise \cite{lt1,lt2}. The best limit so far is given by
\begin{equation}
m_\gamma < 10^{-57}gm\label{e1}
\end{equation}
that is, the photon mass would be very small indeed!\\
It may also be mentioned that there has been a more radical view that the photon itself is superfluous \cite{lt3,marshall,arm,sachs}.\\
The author himself has argued very much on the lines of de Broglie, that the photon has a small mass given by
\begin{equation}
m_\gamma = 10^{-65}gm\label{e2}
\end{equation}
This conclusion follows from a Planck scale underpinning model and is compatible with special relativity \cite{bhtd,mp,fpl,uof}. Curiously enough, from the completely different point of view of Thermodynamics, Landsberg had shown that the mass in (\ref{e2}) is the smallest possible mass in the universe \cite{land}. In any case (\ref{e2}) is compatible with (\ref{e1}).\\
It is interesting to note that the mass of the photon given in (\ref{e2}) has some experimental support. This mass would imply a dispersive effect in High Energy Cosmic Rays which we receive from deep outer space - and it appears that we may already be observing such effects \cite{mp,pav}.\\
This apart another experimental evidence for the photon mass (\ref{e2}) comes from laboratory diffraction experiments \cite{vigier}.\\
Finally it may be mentioned that this photon mass would imply that the Coulomb potential becomes a short range Yukawa potential with a range $\sim 10^{28}cms$, that is the radius of the universe itself. Such a Yukawa potential would lead to a small shift in the hyperfine energy levels as shown elsewhere \cite{ijmpe}.
\section{Further Theoretical Support}
We now give a further theoretical justification for the above. We first observe that as is well known \cite{neeman}, Maxwell's equations can be written in the following form
\begin{equation}
{\bf \Psi} = \vec{E} + \imath \vec{B},\label{e3}
\end{equation}
\begin{equation}
\vec{\nabla} \times {\bf \Psi} = \imath \dot{{\bf \Psi}}\label{e4}
\end{equation}
\begin{equation}
\vec{\nabla} \cdot {\bf \Psi} = 0\label{e5}
\end{equation}
Equations (\ref{e3}) to (\ref{e5}) will be useful in the sequel.\\
We next observe that Maxwell's equations have been deduced in a fashion very similar to the Dirac equation, from first principles \cite{gersten}. In this deduction, we use the usual energy momentum relation for the photon
$$E^2 - p^2 c^2 = 0$$
and introduce matrices given in
$$S_x = \left(\begin{array}{ll}
0 \quad 0 \quad 0\\
0 \quad 0 \quad -\imath\\
0 \quad \imath \quad 0
\end{array}\right)\, , S_y = \left(\begin{array}{ll}
0 \quad 0 \quad \imath\\
0 \quad 0 \quad 0\\
-\imath \quad 0 \quad 0
\end{array}\right)\, ,$$
\begin{equation}
S_z = \left(\begin{array}{ll}
0 \quad -\imath \quad 0\\
\imath \quad 0 \quad 0\\
0 \quad 0 \quad 0
\end{array}\right)\, , I^{(3)} = \left(\begin{array}{ll}
1 \quad 0 \quad 0\\
0 \quad 1 \quad 0\\
0 \quad 0 \quad 1
\end{array}\right)\, ,\label{e6}
\end{equation}    
from which we get
$$\left(\frac{E^2}{c^2} - {\bf p^2}\right) {\bf \Psi} = \left(\frac{E}{c}I^{(3)} + {\bf p \cdot S}\right){\bf \Psi}$$
\begin{equation}
- \left(\begin{array}{ll}
p_x\\
p_y\\
p_z
\end{array}\right) \, ({\bf p \cdot \Psi}) = 0,\label{e7}
\end{equation}
where $\Psi$ is a three component wave function and in general bold letters denote vector quantization.\\
Equation (\ref{e7}) implies
\begin{equation}
\left(\frac{E}{c} I^{(3)} + {\bf p \cdot S}\right) {\bf \Psi} = 0,\label{e8}
\end{equation}
\begin{equation}
{\bf p \cdot \Psi} = 0,\label{e9}
\end{equation}
where $S$ is given in (\ref{e6}). There is also an equation for ${\bf \Psi^*}$ namely
\begin{equation}
\left(\frac{E}{c}I^{(3)} - {\bf p \cdot S}\right) {\bf \Psi^*} = 0,\label{e10}
\end{equation}
\begin{equation}
{\bf p \cdot \Psi^*} = 0,\label{e11}
\end{equation}
It is then easy to verify (Cf.ref.\cite{gersten}) that with the substitution of the usual Quantum Mechanical energy momentum operators, we recover equations (\ref{e3}) to (\ref{e5}) for ${\bf \Psi}$ and its complex conjugate.\\
Recently a similar analysis has lead to the same conclusion. In fact it has been shown that under a Lorentz boost \cite{dvg},
\begin{equation}
\left(\begin{array}{ll}
\Psi'\\
\Psi^{*'}\end{array}\right) = \left(\begin{array}{ll}
1 - \frac{({\bf S \cdot p})}{mc} + \frac{({\bf S \cdot p})^2}{m(E+mc^2)} \quad \quad 0\\
0 \quad \quad \quad \quad 1 + \frac{({\bf S \cdot p})}{mc} + \frac{({\bf S \cdot p})^2}{m(E+mc^2)}\end{array}\right) \left(\begin{array}{ll}
\Psi\\
\Psi^* \end{array}\right)\label{e12}
\end{equation}
We would like to point out that equations (\ref{e4}), (\ref{e5}), (\ref{e8}) to (\ref{e12}) display the symmetry 
$${\bf p} \to -{\bf p} \quad , \Psi \to \Psi^*$$
We now invoke the Weinberg-Tucker-Hammer formalism (Cf.\cite{dvg}) which gives, for a Lorentz boost equations
\begin{equation}
\phi'_R = \left\{ 1 + \frac{{\bf S \cdot p}}{m} + {({\bf S \cdot p})^2}{m(E + m)}\right\} \phi_R,\label{e13}
\end{equation}
\begin{equation}
\phi'_L = \left\{ 1 - \frac{{\bf S \cdot p}}{m} + {({\bf S \cdot p})^2}{m(E + m)}\right\} \phi_L,\label{e14}
\end{equation}
where the subscripts $R$ and $L$ refer to the states of opposite helicity, that is left and right polarised light in our case.\\
We now observe that equations (\ref{e12}) and (\ref{e13})-(\ref{e14}) are identical, but there is a curious feature in both of these, that is that the photon of electromagnetism is now seen to have a mass $m$.
\section{Discussion}
It is interesting to note that it has been demonstrated that the mass of the photon is incompatible with the magnetic monopole \cite{ignat}. Indeed the author himself has presented different arguments to the effect that there are no magnetic monopoles \cite{nc}. It may be mentioned that Dirac the originator of the idea of the magnetic monopole, himself expressed his pessimism about the existence of the magnetic monopole as long back as 1981, during the fiftieth year of the monopole seminar \cite{dirac}.\\
Returning to the mass of the photon, it can be argued that this is a result of the non commutativity of spacetime at a micro scale. We observe that a photon mass would imply the equation
\begin{equation}
\partial^{\mu} F_{\mu \nu} = - m^2 A_{\nu},\label{e15}
\end{equation}
where we have the usual equations of electromagnetism
\begin{equation}
\partial^{\mu} A_{\mu} = 0,\label{e16}
\end{equation}
\begin{equation}
F_{\mu \nu} = \partial_{\mu} A_{\nu} - \partial_{\nu} A_{\mu}\label{e17}
\end{equation}
We note that from (\ref{e17}) we get
\begin{equation}
\partial^{\mu} F_{\mu \nu} = D A_{\nu} - \partial^{\mu} \partial_{\nu} A_{\mu}\label{e18}
\end{equation}
where $D$ denotes the D'Alembertian. In view of (\ref{e16}), the second term on the right side of (\ref{e18}) would vanish, provided the derivates commute. In this case we would return to the usual Maxwell equations. However in the non commutative case this extra term is
$$p^2 A_\mu \sim m^2 A_\mu$$
remembering that we are in units $c = 1 = \hbar$.\\
Thus because of the non commutativity we get (\ref{e15}) instead, of the usual Maxwell equation.\\
The question of non commutativity and mass generation in the context of gauge theory has been studied by the author elsewhere (Cf.\cite{annales,ijmpe2,uof}).\\
In any case the above points to the fact that there would be no massless particles in nature. The point is, that in an idealized situation in which the radius $R \to \infty$, the mass $m_\gamma \to 0$.

\vspace{5 mm}
\begin{flushleft}
{{\bf \large {APPENDIX}}}
\end{flushleft}
{\bf 1. Non Commutativity and Mass Generation:}\\ \\
Our starting point is the Dirac equation (using natural units $c = 1 = \hbar$),
\begin{equation}
\left(\gamma^\mu p_\mu\right) \psi = 0\label{ea1}
\end{equation}
Remembering that the operator in (\ref{ea1}) is
$$\gamma^\circ p^\circ - \vec{\gamma}_\circ \vec{p},$$
we multiply on the left side by
$$\gamma^\circ p^\circ + \vec{\gamma}_\circ \vec{p}$$
This gives us
\begin{equation}
\left[\left(p^2_0 - \vec{p}^2\right) - \imath \vec{\sum}\cdot \left(\vec{p}\times \vec{p}\right) + \gamma^\imath \gamma^\circ B_{\imath \circ}\right]\psi = 0\label{ea2}
\end{equation}
where $\vec{\sum} = \left(\begin{array}{ll}
\vec{\sigma} \quad 0 \\
0 \quad \vec{\sigma}
\end{array}
\right)$. In (\ref{ea2}), the first term is the usual energy momentum term which leads to the massless Klein-Gordon or D'Alembertian operator. The second term is well known in a non relativistic approximation with a small external magnetic field that is switched on, this term leads to a spin orbit coupling. It is the third and last term that is the extra effect due to the noncommutative character of spacetime, that is due to the fact
\begin{equation}
B_{\mu \nu} = \left[p_\mu , p_\nu \right] \ne 0\label{ea3}
\end{equation}
We identify this extra term with the mass term, viz.,
\begin{equation}
\gamma^\imath \gamma_\circ B_{\imath \circ} = m^2\label{ea4}
\end{equation}
We will justify this identification in a moment. With this identification, and in the absence of an external magnetic field, in which case the second term in (\ref{ea2}) disappears, (\ref{ea2}) goes over to the Klein-Gordon equation for a massive particle.\\
Infact it has already been discussed in detail that in the above noncommutative case, (\ref{ea3}),
\begin{equation}
B_{\mu \nu} = p_\mu p_\nu - p_\nu p_\mu = \partial_\mu \partial_\nu - \partial_\nu \partial_\mu = e F_{\mu \nu}\label{ea5}
\end{equation}
where $F_{\mu \nu}$ is the usual electromagnetic field tensor. (The deep relation of (\ref{ea5}) with the Weyl gauge invariant electromagnetic potential has also been discussed in detail in the above references). Because of (\ref{ea5}) and because of the fact that,
$$
\gamma^\imath \gamma^\circ = \alpha^\imath$$
where $\vec{\alpha}$ denotes the velocity operator, at the Compton wavelength where momentum is $m (= mc)$, the extra term becomes
$$\frac{e^2}{l^2} \sim m^2$$
in agreement with (\ref{ea4}) due to the definition of the Compton length as the electron radius viz.,
$$l \sim e^2/m \sim \frac{1}{m}$$
The massive Klein-Gordon equation then, in the usual formulation leads back to the Dirac equation (\ref{ea1}) but this time with the usual mass term.\\ \\
{\bf 2. An Experimental Test for the Photon Mass:}\\ \\
It is well known that for a massive vector field interacting with a magnetic dipole of moment ${\bf M}$, for example the earth itself, we would have with the usual notation 
$${\bf A}(x) = \frac{\imath}{2} \int \frac{d^3 k}{(2\pi )^3}{\bf M \times k} \frac{e^{\imath k, x}}{{\bf k}^2 + \mu^2} = - {\bf M \times \nabla} \left(\frac{e^{-\mu r}}{8\pi r}\right)$$
\begin{equation}
{\bf B} = \frac{e^{-\mu r}}{8 \pi r^3} | {\bf M}| \left\{\left[ \hat{r} (\hat{r} \cdot \hat{z}) - \frac{1}{3} \hat{z}\right] (\mu^2 r^2 + 3\mu r + 3) - \frac{2}{3} \hat{z} \mu^2 r^2\right\}\label{e4b}
\end{equation}
Considerations like this have yielded in the past an upper limit for the photon mass, for instance $10^{-48}gms$ and $10^{-57}gms$. Nevertheless (\ref{e4b}) can be used for a precise determination of the photon mass. It may be mentioned here that contrary to popular belief, there is no experimental evidence to indicate that the photon mass is zero!

\end{document}